\pdfoutput=1
\documentclass[twocolumn,amsmath,amssymb,amsthm,PRL]{revtex4}
\usepackage[latin1]{inputenc}
\usepackage{mathrsfs}
\usepackage{graphicx,epsf,subfigure,pstricks}
\usepackage{color}

\begin{document}

\author{
J\'er\'emy Hure
}
\author{
Beno\^it Roman
}
\author{Jos\'e Bico
}

\title{Stamping and wrinkling of elastic plates}

\affiliation{PMMH, CNRS UMR 7636, UPMC \& Univ. Paris Diderot, ESPCI-ParisTech, 10 rue Vauquelin, 75231 Paris Cedex 05, France.}

\begin{abstract}
We study the peculiar wrinkling pattern of an elastic plate stamped into a spherical mold. 
We show that the wavelength of the wrinkles decreases with their amplitude, but reaches a maximum when the amplitude is of the order of the thickness of the plate. 
The force required for compressing the wrinkled plate presents a maximum independent of the thickness. 
A model is derived and verified experimentally for a simple one-dimensional case. 
This model is extended to the initial situation through an effective Young modulus 
representing the mechanical behavior of wrinkled state.
The theoretical predictions are shown to be in good agreement with the experiments.
This approach provides a complement to  the "tension field theory" developed for wrinkles with unconstrained amplitude.
\end{abstract}

\maketitle 

\begin{figure}[!h]
\begin{center}
\subfigure[]{\includegraphics[height=2.8cm]{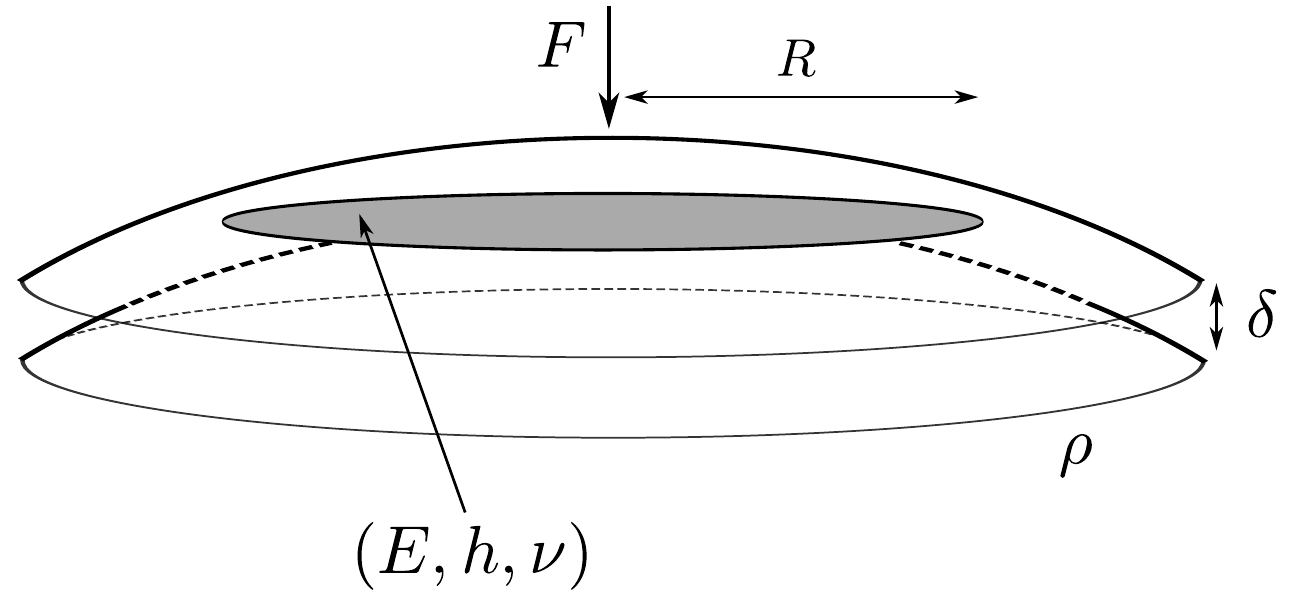}}
\begin{tabular}{cc}
\subfigure[]{\includegraphics[height=3.5cm]{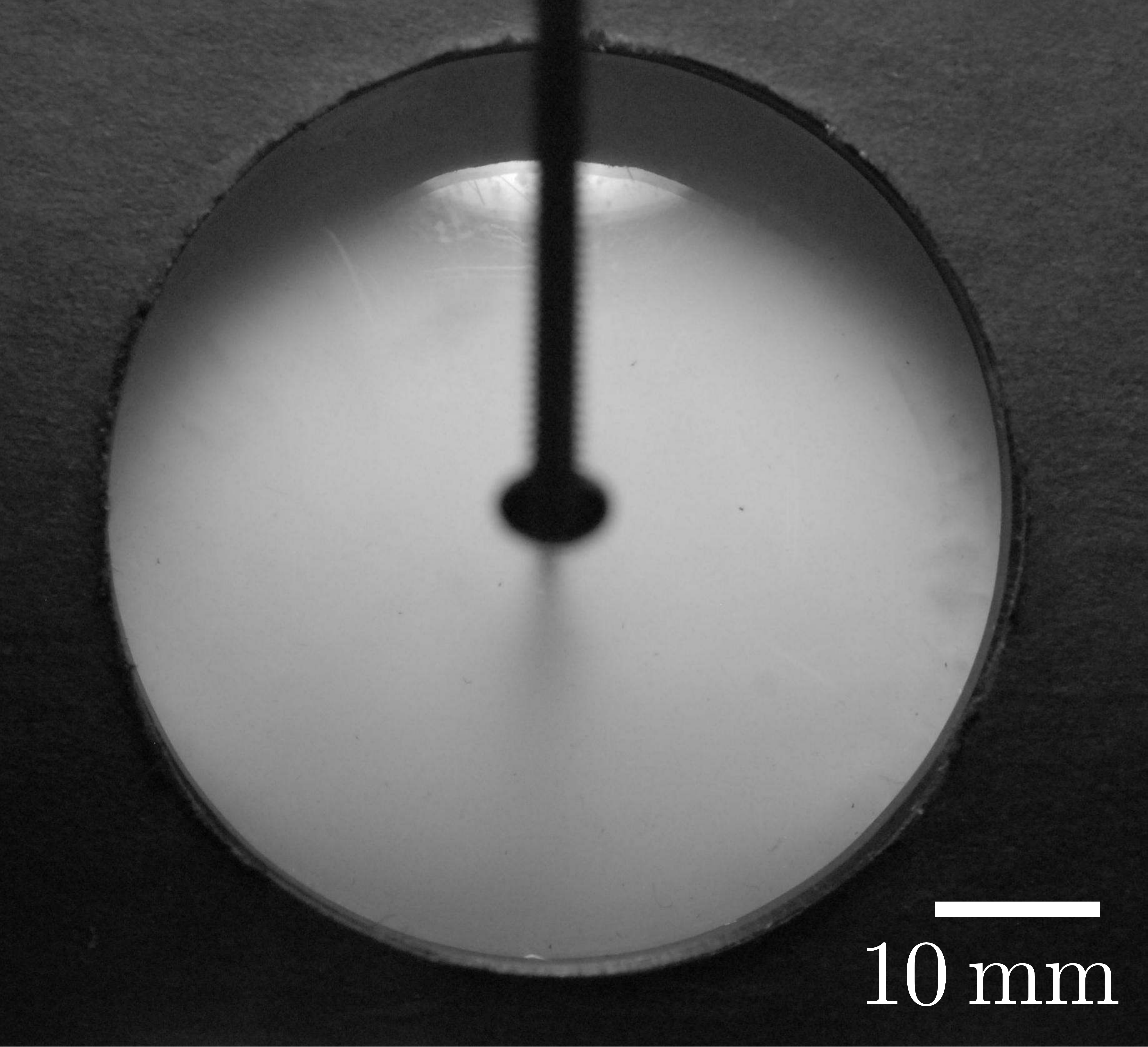}} &
\subfigure[]{\includegraphics[height=3.5cm]{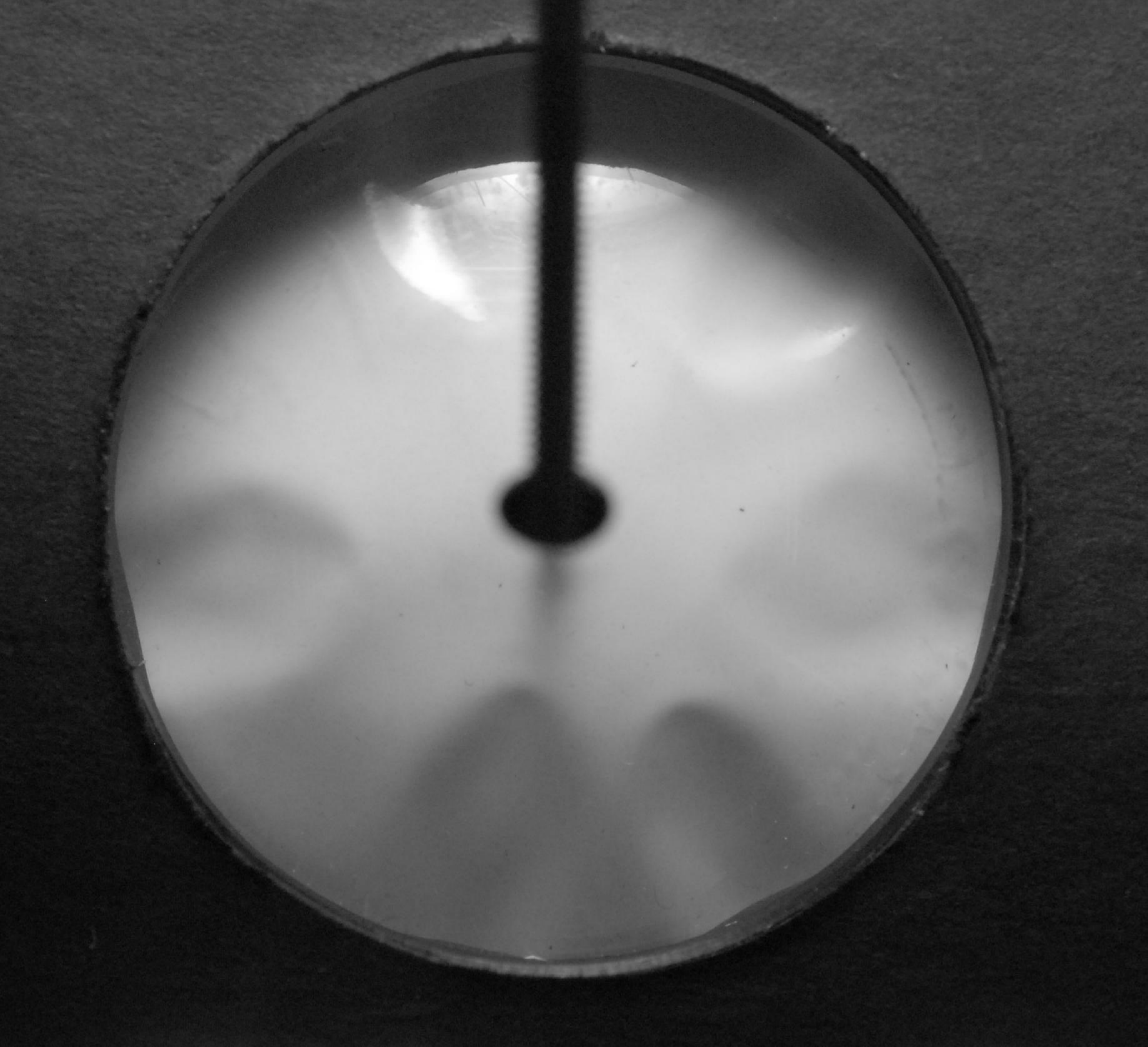}} \\
\subfigure[]{\includegraphics[height=3.5cm]{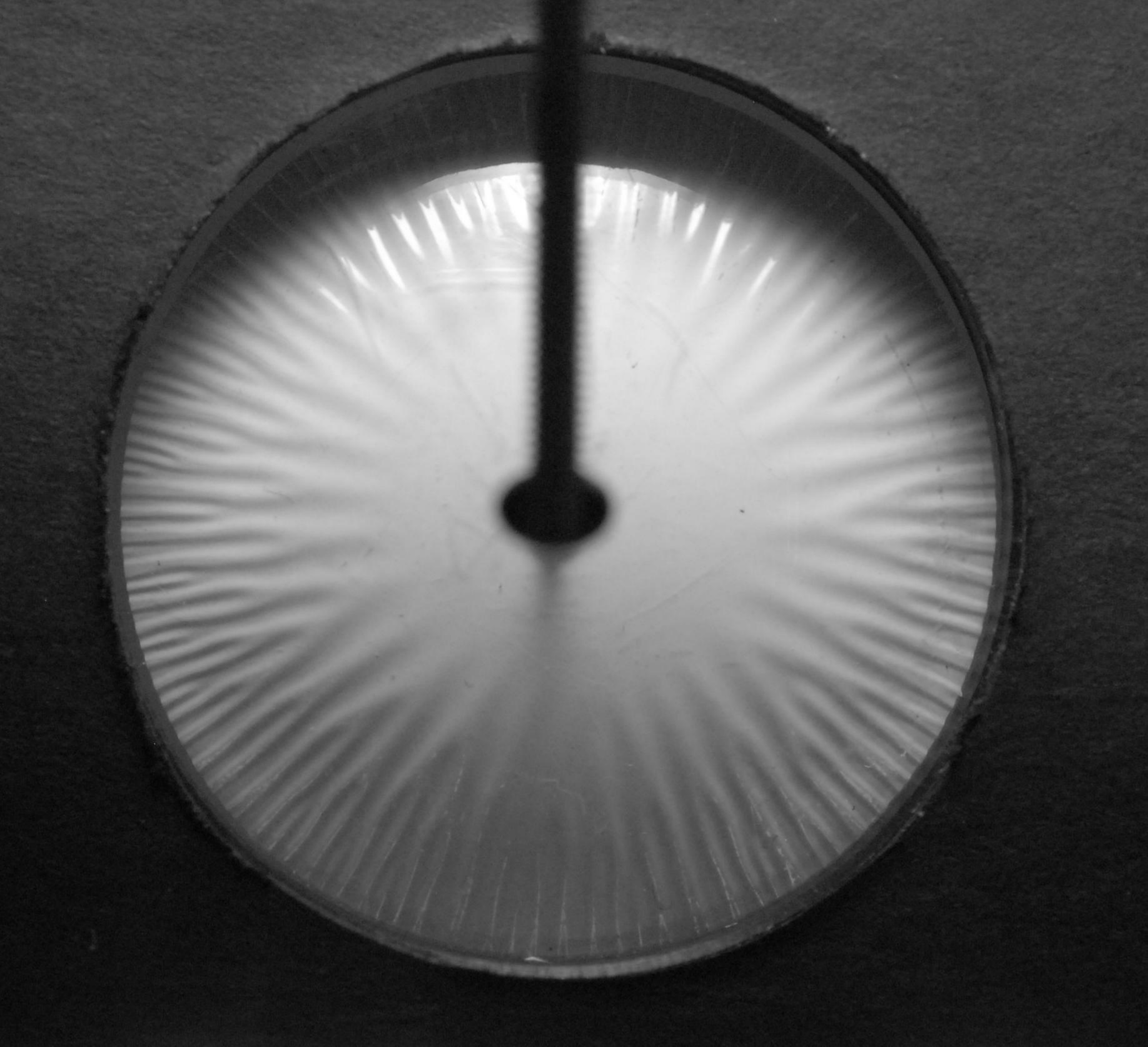}} & \subfigure[]{\includegraphics[height=3.5cm]{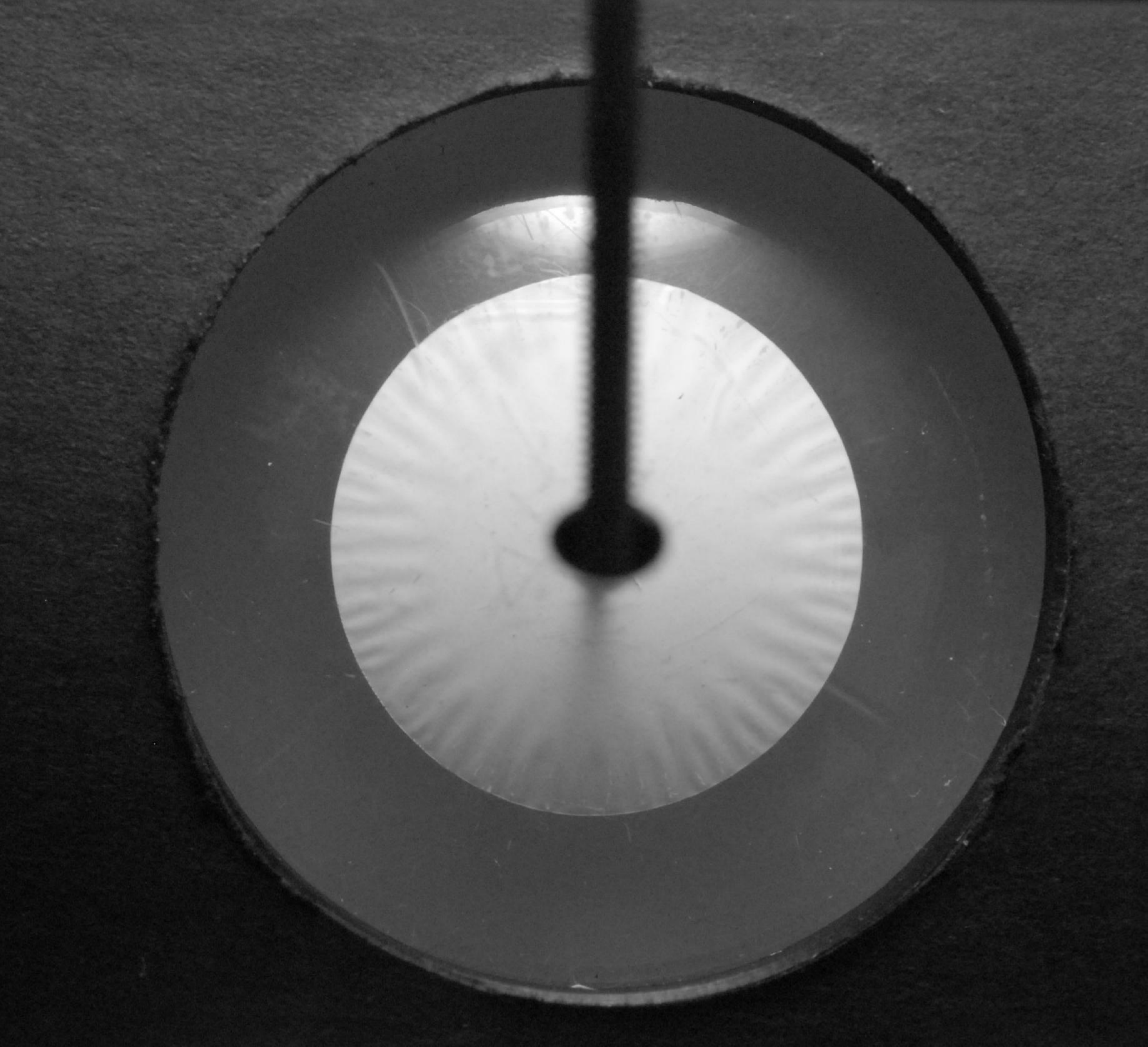}}
\end{tabular}
\end{center}
\vspace{-5mm}
\caption{Top: Experimental setup. A circular plate of radius $R$ is compressed between two rigid transparent hemispherical dies of radius $\rho$. The gap $\delta$ between the spheres is imposed. Bottom: Experimental observations of wrinkles formed when a plate is compressed between two hemispherical dies. $\delta$ is decreased from (b) to (d) while $R$ is maintained constant. (d) and (e) are characterized by the same $\delta$, but different values for $R$ ($\rho = 58\,$mm, $h=58\,\mu$m and $R = 12.5/25\,$mm).}
\label{experiment}
\end{figure}

Wrinkling patterns are observed when thin plates are put under compression, spanning scales from geological patterns \cite{porada}, skin wrinkles resulting from aging processes or scars \cite{cerdawrin,cerdascar} to cells locomotion generating strains on substrates \cite{harris}.
They have important applications in micro-engineering such as the formation of controlled patterns \cite{bowden98}, or the estimation of  mechanical properties from the number and the extent of the wrinkles \cite{genzer,geminard, huang07, stafford04}.
While most studies have focused on near threshold patterns, recent contributions have pushed further 
the description of finely wrinkled plates, \textit{i.e.}, well above the initial buckling threshold \cite{bennypnas,king12}. 
These studies consider plates submitted to strong in-plane tension on the boundaries, which prevents stress focusing commonly observed in crumpled paper~\cite{witten2007}. 
In these descriptions, wrinkles are assumed to totally relax compressive stresses, as in traditional tension field theory \cite{nasa,mansfield}.
In addition, both amplitude and wavelength of the wrinkles vanish with the thickness of the plate. 
We propose to study a conceptually different wrinkling regime where the boundaries are free from tension but the amplitude of the wrinkles is highly constrained. We focus on the simplest example, an elastic plate compressed in a spherical mold with a confinement defined by a gap $\delta$ (Fig.~\ref{experiment}). 
This stamping configuration is common in industrial processes where metal plates are plastically embossed, the mismatch in Gaussian curvature generally leading to regular wrinkles \cite{yu85,stronge86, forming}. 
In the elastic case, crumpling singularities first appear as the mold is progressively closed down (Fig.~\ref{experiment}c) and evolve into a pattern  of
apparently smooth radial wrinkles for high confinement (Fig.~\ref{experiment}d-e).
We show that constraining the amplitude does not lead to the collapse of compressive stress. 
Instead, the wrinkling pattern derives from a nontrivial balance between compression and bending stresses, with surprising consequences : the  wavelength of the wrinkles does not vanish and the constraining force reaches a maximum independent of the thickness of the plate.

\begin{figure}
\begin{center}
\includegraphics[height=3.2cm]{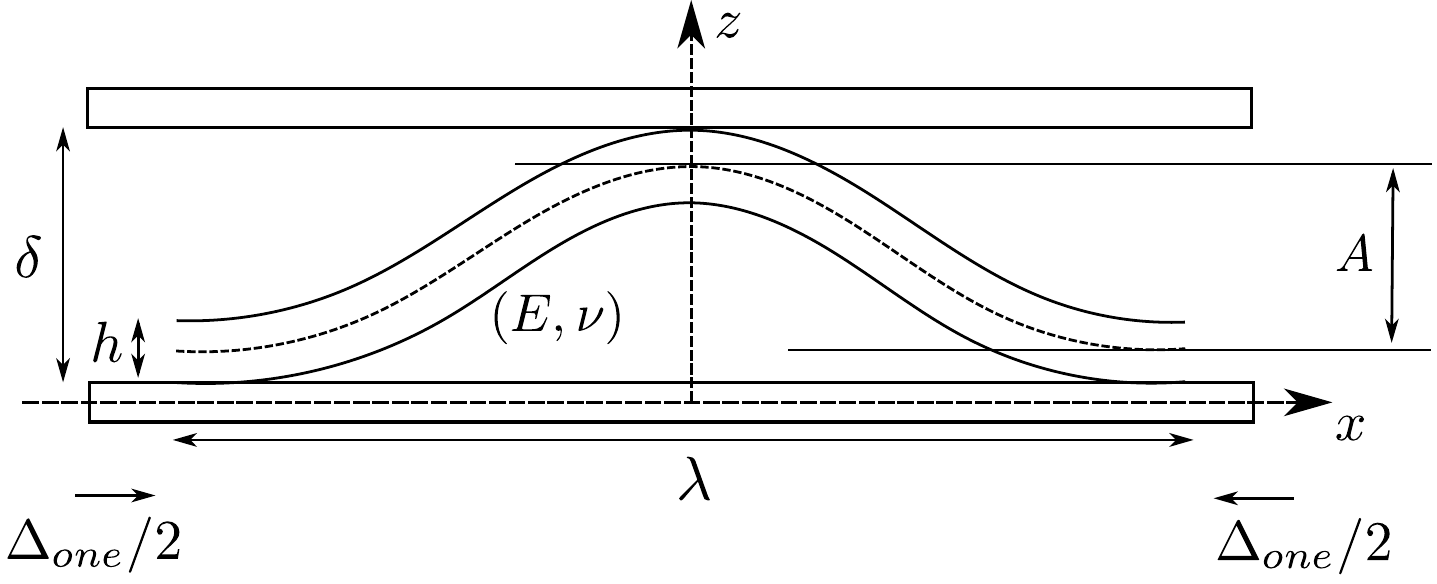}
\end{center}
\vspace{-5mm}
\caption{Experimental setup: a plate of length $L$, thickness $h$ and unit width is subjected to symmetrical displacements at the ends $u_x(\pm L/2) = \pm \Delta/2$ and a maximal deflection $\delta$. We consider that $n$ similar wrinkles are formed ($\Delta_{one}=\Delta/n$) of wavelength $\lambda=L/n$ and amplitude $A$, with $A = \delta - h$. }
\label{sketch1D}
\end{figure}

\emph{One-dimensional problem.} We start by considering the simpler problem of a  plate of length $L$, thickness $h$ and unit width. In-plane displacements are imposed at both ends $u_x(\pm L/2) = \pm \Delta/2$ (Fig.~\ref{sketch1D}). 
In addition, the out-of-plane displacement of the plate is constrained to a maximal value $\delta$. 
The amplitude $A$ of the median plane of the plate thus corresponds to $A = \delta - h$. 
For large values of $\delta$ a single wrinkle forms as the axial force exceeds Euler's critical load \cite{timostab}.
Within small slope approximation, the out-of-plane deflection of a single wrinkle (Fig.~\ref{sketch1D}), is taken as $w(x) = A/2 [1+\cos{(2\pi x/\lambda)}]$, where $\lambda$ is the wavelength. 
The associated bending energy thus reads:
\begin{equation}
\mathcal{E}_b = \frac{B}{2} \int_{-\lambda/2}^{\lambda/2} \left( \frac{\partial^2 w}{\partial x^2} \right)^2 = \frac{B \pi^4 A^2}{\lambda^3}
\label{eq1}
\end{equation}
with $B=Eh^3/[12(1-\nu^2)]$ the bending modulus of the plate ($E$ and $\nu$  the material Young's modulus and Poisson's ratio). 
We are here interested in the non-classical limit $\delta \to h$ where in-plane stress $\sigma$ and strain $\epsilon$ cannot be neglected.
According to the in-plane equilibrium equation $\partial_x\sigma = 0$, $\sigma$ and $\epsilon$ are constant along the plate, leading to $\epsilon = \int_{-\lambda/2}^{\lambda/2}[\partial u/\partial x + (1/2)(\partial w /\partial x)^2] \,dx / \lambda$. 
The corresponding stretching energy can be written as:
\begin{equation}
\mathcal{E}_s = \frac{S}{2} \epsilon^2 \lambda= \frac{S}{2} \left(\frac{\pi^2 A^2}{4 \lambda^2} - \frac{\Delta_{one}}{\lambda}\right)^2 \lambda
\label{eq2}
\end{equation}
where $\Delta_{one}$ corresponds to the in-plane displacement for a wrinkle and $S=Eh/(1-\nu^2)$ is the stretching modulus of the plate \footnote{Here we consider that the deformation is zero along the transverse direction, thus Hooke's law gives $\sigma = [E/(1-\nu^2)] \epsilon$}.

We consider a series of $n$ successive sinusoidal wrinkles along the length $L$ of the plate. Such model does not reproduce the continuous evolution of the confined plate, as flat parts are actually observed \cite{roman99,roman02}. Nevertheless, it describes exactly particular shapes taken by the plate all along the stamping process. To simplify the description, we thus propose to join these particular states continuously by considering non-integer values of $n$.
The global energy corresponding to $n$ successive wrinkles is obtained by using the conditions $n\lambda = L$ and $n \Delta_{one} = \Delta$ in equations~(\ref{eq1}) and~(\ref{eq2}):
\begin{equation}
\mathcal{E}_{tot} =   \left[ n^4\frac{B \pi^4 A^2}{L^4} + \frac{S}{2} \left(\frac{n^2 \pi^2 A^2}{4 L^2} - \frac{\Delta}{L} \right)^2 \right] L
\label{eq3}
\end{equation}

\begin{figure}
\begin{center}
\includegraphics[height=5.2cm]{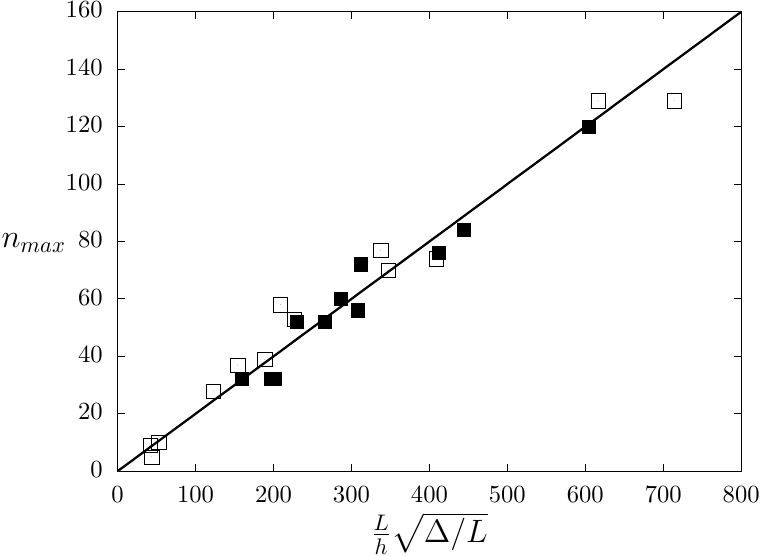}
\end{center}
\vspace{-5mm}
\caption{Maximal number of wrinkles for polypropylene films of thicknesses ranging from $15\,\mu$m to $250\,\mu$m. Open symbols, 1D experiments: strips of length $250\,$mm, width $25\,$mm with $\Delta/L=0.2\%$ and $0.3\%$. Filled symbols, 2D experiments: discs of radius $R$ ranging from $10\,$mm to $70\,$mm, radius of spheres $\rho=58,\,600\,$mm and $n_{max}$ the number of wrinkles at the edge of the discs. For a comparison with 1D experiments, we take $L=2 \pi R$ and $\Delta/L = R^2/8\rho^2$ \cite{majidi08}. The solid line corresponds to the best fit of the data: $n_{max}=0.20\, ( L/h) \sqrt{\Delta/L}$.}
\label{manip1}
\end{figure}

Minimizing the total energy for an imposed amplitude with respect to the number of wrinkles leads to:
\begin{equation}
{n}^{2}=\frac{12\,(\Delta/L)\,{L}^{2}}{3\,{\pi}^{2}\,{A}^{2}+8\,{\pi}^{2}\,h^2}
\label{eqn}
\end{equation}
Counter-intuitively the number of wrinkles tends towards a finite value $n_{max} \simeq 0.39 (L/h)\sqrt{\Delta/L}$ as the amplitude $A$ vanishes \footnote{In the standard inextensible approximation where the stretching energy is neglected, the number of wrinkles would diverge as  $n=2(\Delta L)^{1/2}/\pi A$.}. 
In order to validate this prediction, polypropylene films ($E=2200\,$MPa and $\nu=0.4$, Innovia Films) of thicknesses $h$ ranging from $15\,\mu$m to $250\,\mu$m are cut in bands of length $L=250\,$mm and $25\,$mm of width. The strip is clamped on a rigid plate with an imposed displacement $\Delta$. The resulting blister is confined as illustrated in Fig.~\ref{sketch1D}. Note here that to avoid plastic events, all the experiments are such that $\Delta/L \leq \epsilon_Y$, where $\epsilon_Y \approx 2\%$ is the yield point of the material.
We observe that for large confinement the number of wrinkles saturates to a maximum value in good agreement  with the theoretical power law (Fig.~\ref{manip1}).  
Nevertheless the value of the prefactor is significantly lower than predicted (0.2 instead of 0.39).
We interpret this discrepancy as a consequence of the discrete transition between successive modes  \cite{roman99}.
Indeed the buckling of $n$ wrinkles can result into $n+1$ to $3n$ wrinkles (if all wrinkles split simultaneously), which leads to an ambiguity in the actual mode number.

\begin{figure}
\begin{center}
\includegraphics[height=5.2cm]{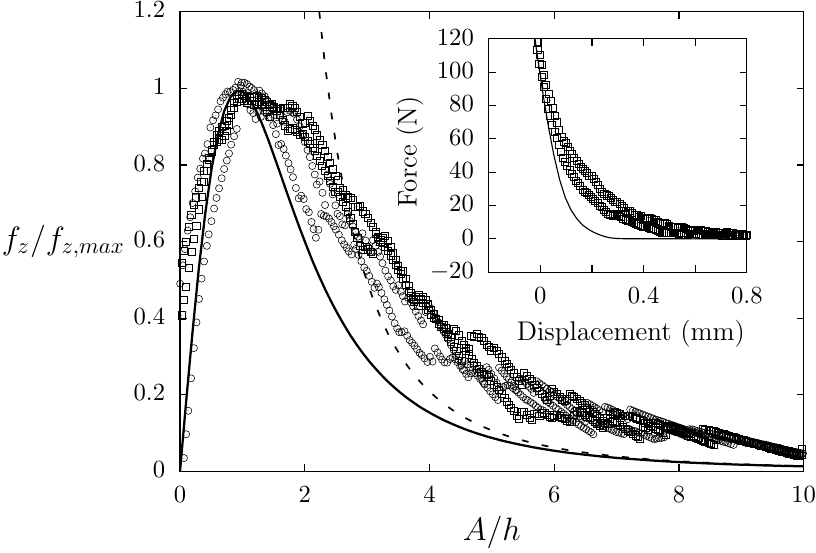}
\end{center}
\vspace{-5mm}
\caption{Evolution of the vertical force (per unit width) $f_z$ required to confine the plate as a function of the dimensionless amplitude $A/h$, defined as $f_z(A/h = 0.94) = f_{z,max}$. The solid line corresponds to eq.~\ref{eq:Fz}, the dashed line to the inextensible approximation. Experiments with strips of length $250\,$mm, width $25\,$mm with $\Delta/L=0.07\%$ (circles) and $0.2\%$ (squares). Inset: Raw force data for $h = 90\,\mu$m, solid line corresponds to $\Delta/L=0$, squares to $\Delta/L = 0.2\%$.}
\label{force}
\end{figure}

In addition to the number of wrinkles, the load (per unit width) required for compressing the plate can finally be derived by differentiating the elastic energy:
\begin{subequations}
\begin{equation}
f_x = \frac{\partial \mathcal{E}_{tot}}{\partial \Delta} = \frac{(\Delta/L)}{(1 + 3\alpha^2/8)}\,S
\label{eq:Fx}
\end{equation}
\begin{equation}
f_z = \left| \frac{\partial \mathcal{E}_{tot}}{\partial A} \right| = \frac{24\,\alpha\,(L/h)\,(\Delta/L)^2}{9\,{\alpha}^{4}+48\,{\alpha}^{2}+64}\,S 
\label{eq:Fz}
\end{equation}
\end{subequations}
where $\alpha = A/h$ is the relative amplitude.
Within the limit $A \gg h$, both forces are proportional to the bending stiffness $Eh^3$ (the plate can then be considered as inextensible) and increase with the confinement. 
However the vertical load vanishes for high compression ($\alpha \ll 1$) after reaching a maximum $f_{z,max} \simeq 0.24 EL(\Delta/L)^2$ (with $\nu=0.4$) for  $A/h \simeq 0.94$, while the axial force (per unit width) tends towards $(\Delta/L)S$. When $A = 0$, we recover the simple case of a flat plate under lateral compression where buckling is inhibited, and $f_z$ vanishes. The experimental measurement of the vertical force 
 is delicate since any slight misalignment generates an additional force during the compression.  
Capturing  the limit $A \rightarrow 0$ also requires an accurate positioning of the rigid dies.
To limit artifacts due to misalignment, the force due to the wrinkles $f_z$ is obtained by substracting the force given by the tensile machine for $\Delta/L = 0$ (corresponding to a flat plate) to the one for $\Delta/L \neq 0$. 
As shown in the inset of Fig.~\ref{force}, this difference clearly indicates the presence of a maximum, as predicted by our description. 
The precise localization of $A = 0$ is still difficult and is inferred from the position corresponding to the maximum force,  $f_z(A/h = 0.94)=f_{z,max}$ (Fig.~\ref{force}).
We observe that for a fixed value of $A$, the axial force in eq.~(\ref{eq:Fx}) is proportional to the applied strain $\Delta/L,$ which brings us to define an effective Young's modulus:
\begin{equation}
\frac{f_x}{h} = \left[ \frac{E_{eff}}{1-\nu^2} \right] \frac{\Delta}{L} \Rightarrow E_{eff} = \frac{E}{1 + \frac{3}{8} \left({A}/{h}\right)^2}.
\end{equation}
This effective modulus increases from zero to $E$ as the plate is progressively confined.

\begin{figure}[h!]
\begin{center}
\includegraphics[height=5.2cm]{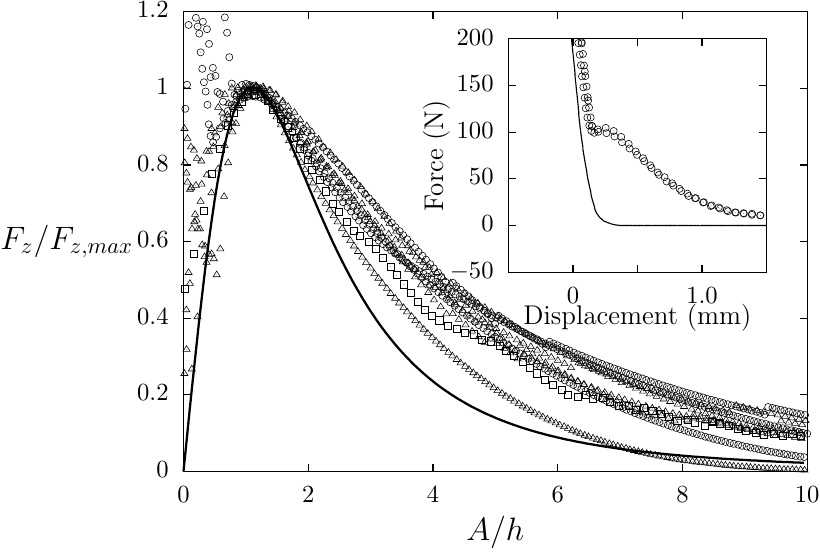}
\end{center}
\vspace{-5mm}

\caption{Evolution of the vertical force $F_{z1}$ required to confine the plate as a function of the dimensionless amplitude $A/h$, defined as $F_{z1}(A/h = 1.14) = F_{z1,max}$. Experiments with discs of radius $25\,$mm, thicknesses $h=90\,\mu$m (squares), $150\,\mu$m (circles) and $250\,\mu$m (triangles), with $\rho=58\,$mm. The solid line corresponds to the theoretical prediction. Inset: Raw force data, the solid thin line corresponds stamping without plate, circles to $h=150\,\mu$m.}
\label{2D}
\end{figure}

\emph{Two dimensional patterns.} 
We consider elastic discs (same material) with radius $R$ (from $10\,$mm to $70\,$mm)  and thickness (from $30\,\mu$m to $250\,\mu$m) embossed between hemispherical stamps of radius $\rho$ ($58\,$mm and $600\,$mm), separated by a distance $\delta$ (Fig.~\ref{experiment}). The compaction force is measured as the difference of the raw force with the force obtained without the plate (Fig.~\ref{2D}).
The orthoradial compression responsible for the formation of wrinkles is a consequence of Gauss {\it Theorema Egregium}: wrapping a sphere with a planar sheet implies stretching or compression in addition to bending \cite{struik}.
In terms of scaling, the perfect contact between the plate and the stamp involves a typical strain $\epsilon \sim (R/\rho)^2$ and a stretching energy $EhR^6/\rho^4$ \cite{majidi08,hure11}.
The ratio of the stretching energy to the typical bending energy $Eh^3R^2/\rho^2$ thus yields the dimensionless parameter $R/\sqrt{\rho h}$.
Therefore forming wrinkles is not expected to release energy in the case $R^2 \ll {\rho h}$, {\it i.e.} when the bending is dominant over stretching energy.
In the experiment, if $R$ is small enough, the plate indeed remains unwrinkled when the mold is closed. 

We focus on the opposite limit $R^2 \gg {\rho h}$, especially for $\delta \simeq h$, where hierarchical wrinkles appear beyond a certain distance from the center of the plate.  
The minute amplitude of the wrinkles lead us to assume that the wrinkles only involve smooth deformations as a perturbation from the state $\delta = h$, where the plate lays unwrinkled on the spherical cap. 
In other words, we conjecture~\footnote{To show that a defocusing transition takes place~\cite{roman12} we would need to exhibit a smooth solution where bending and stretching energies are balanced. Our approximate 2D solution is not  completely convincing because it does not consider the matching condition between wrinkles with continuously varying wavelength} that the observed transition in fig.~\ref{experiment}c,d corresponds to stress defocusing similar to the transition described in~\cite{roman12}.
Within this assumption, the scaling law for the maximal number of wrinkles derived in the 1D situation can be adapted to the circular plate. 
Since the effective orthoradial strain is given by $(r/\rho)^2$, the number of wrinkles observed  at the edge of the plate for high confinement is expected to follow $n_{max} \sim R^2/\rho h$. 
To compare quantitatively 1D and 2D experimental results, we consider that $L=2\pi R$ and $\Delta/L = R^2/8\rho^2$, which corresponds to the strain at the edge of a plate completely in contact with the sphere. 
Fig.~\ref{manip1} shows a good agreement between the prediction for the wavelength and the experiments (with the same prefactor as in 1D). 
Note that using eq.~\ref{eqn} near the middle of the plate where $\epsilon_{\theta \theta } =0$ would give $n = 0$, which is not the case. The detailed description of the wrinkling cascade can not be captured by a simple 1D model and the transition between flat and wrinkled parts of a plate has previously been shown to be problematic \cite{bennypnas}.  
However our scaling would predict a number of wrinkles at the middle of the plate $n_{max}/4$. 
If the transition were set by period doubling this number would correspond to three generation of wrinkles, which is qualitatively observed (Fig.~\ref{experiment}).

In order to describe the extension of the smooth region and the stamping force, we consider a model plate with an effective Young modulus $E_{eff}$ in the orthoradial direction and $E$ in the radial direction. 
As a first approximation we assume that wrinkles appear for $\sigma_{\theta \theta} < 0$.
We thus consider $E_{eff} = E$ for $\sigma_{\theta \theta}(r)\geq 0$ when no wrinkles are formed and $ E_{eff} ={E}/[{1 + \frac{3}{8} \left({A}/{h}\right)^2}]$ for $\sigma_{\theta \theta}(r) < 0$. 
The constitutive law for this effective anisotropic material can be derived from the initial Hooke's law:
\begin{equation}
\begin{cases}
E_{eff}\,\epsilon_{\theta\theta} & = \sigma_{\theta\theta} \\
E\,\epsilon_{rr}  &= \sigma_{rr}
\end{cases}
\end{equation}
where we take for simplicity the Poisson ratio of the effective plate equal to zero. The corresponding strains are given by $\epsilon_{rr} = \partial u/\partial r + (1/2)(\partial w/\partial r)^2$ and $\epsilon_{\theta \theta} = u/r$, with $u$ and $w$ are the radial displacement and the deflection of the plate, respectively. In addition we assume the effective stress field in the finely wrinkled region is axisymmetric $\sigma_{r\theta} = 0$ \cite{bennypnas}. Within the limit of high compression ($\delta \sim h$), $w$ is expected to follow $w(r)=-r^2/2\rho$. 
The different strains and the location $a$ of the transition between the smooth and the wrinkled regions are finally derived by solving the equilibrium equation $\partial \sigma_{rr}/\partial r+(\sigma_{rr}-\sigma_{\theta \theta})/r=0$ with the boundary conditions $u(0)=0$, $\sigma_{rr}(R)=0$, $\sigma_{rr}(a^-)=\sigma_{rr}(a^+)$  and $\sigma_{\theta\theta}(a)=0$. 
The analytic solution provides a description of the confined plate from finite to zero amplitude. 
Although the radial stress is always tensile, the orthoradial stress is only tensile in the central region of the sheet and progressively becomes compressive towards the periphery where wrinkles are observed (Fig.~\ref{experiment}).
In the case of high confinement, the radius of the unwrinkled zone is  $a=R/\sqrt{3}$, in qualitative agreement with our experimental data.

In addition to the description of the compressed plate, the combination of both 1D and 2D approaches also provides a simple estimate of the force required for stamping the plate, which is relevant for practical applications. 
Indeed reducing the amplitude of the wrinkles requires a force $F_{z1}$ that was derived for the 1D situation (Eq.~\ref{eq:Fz}).
By replacing $L$ by $R$ and $\Delta/L$ by $(R^2/\rho^2)$, we thus expect $F_{z1} \propto E R^6/\rho^4$. More precisely, we compute this force as $F_{z1} = \int_a^R f_z\, 2\pi rdr$ (with $\Delta/L = \epsilon_{\theta \theta}$ and $L= 2 \pi r$). Moreover, curving the plate induces an additional pressure $P= h(\sigma_{rr}+\sigma_{\theta \theta})/\rho$ and thus a curving force  $F_{z2}\sim Eh R^4/\rho^3$. 
However, in the relevant limit $ R^2 \gg \rho h$, we expect $F_{z2} \ll F_{z1}$. 
Measuring the actual force is more delicate than in the 1D experiment since obtaining a reference force would require a sphere covered with a spherical shell with the same material properties as the compressed sheet \footnote{Misalignment is here also partially compensated by  substracting from the measured compaction force the reference force obtained in the same configuration but without the disc (Fig.~\ref{2D} inset)}.

Nevertheless the evolution of the stamping force with the dimensionless amplitude exhibits a fair agreement with the prediction (Fig.~\ref{2D}).
In addition, we verified that the maximum of the force follows $F_{z1,max} = [(1.85 \pm 0.05)10^{-3}] ER^6/ \rho^4$. The prefactor given by our model is $4.10^{-3}$. 
The discrepancy is however consistent with the difference between the predicted and observed maximal number of wrinkles in 1D: considering that the number of wrinkles is half the one given by eq.~\ref{eqn} would indeed decrease the estimate of the maximal vertical force $f_z$ by a factor close to 2.

To conclude, we have studied experimentally and theoretically the packing of wrinkled thin plates constrained in amplitude. 
We have broadened the one-dimensionnal analysis  \cite{roman99} for vanishing amplitudes with two novel characteristics: the number of wrinkles tends towards a finite value and the stamping force exhibits a maximum value. 
We have finally shown how these results can be relevant in a 2D geometry by using an effective plate that takes into account the presence of wrinkles. 
The effective compressive Young's modulus of the wrinkled zone which depends on the imposed wrinkle amplitude is analogous to the 
complete collapse of compressive stress in plates with unconstrained amplitude but under large tension. 
Our approach can be applied to other stamping geometries, such as negative curvature shapes where we expect different wrinkling pattern.

We thank Olivier Brouard for his help in designing the experimental setup. This study was partially funded by the ANR project MecaWet. 
\newpage

\bibliographystyle{apsrev4-1}
\bibliography{spherebib}

\end{document}